\documentclass[rog]{agutex}

\usepackage{aas_macros}
\usepackage{graphicx}
\usepackage[normalem]{ulem}
 \setkeys{Gin}{draft=false}

\authorrunninghead{D. BANERJEE AND S. KRISHNA PRASAD}

\titlerunninghead{WAVES IN CORONAL HOLES}

\authoraddr{Corresponding author: D. Banerjee,
 Indian Institute of Astrophysics,Koramangala, Bengaluru 560034, India.
(dipu@iiap.res.in)}

\begin{document}

\title{MHD Waves in Coronal Holes}

 \authors{D. Banerjee\altaffilmark{1},
 S. Krishna Prasad\altaffilmark{2}}

\altaffiltext{1}{Indian Institute of Astrophysics, Koramangala, Bengaluru 560034, India.}
\altaffiltext{2}{Astrophysics Research Centre, School of Mathematics and Physics, Queens University Belfast, Belfast BT7 1NN, UK}

\begin{abstract}
Coronal holes are the dark patches in the solar corona associated with relatively cool, less dense plasma and unipolar fields. The fast component of the solar wind emanates from these regions. Several observations reveal the presence of magnetohydrodynamic (MHD) waves in coronal holes which are believed to play a key role in the acceleration of fast solar wind. The recent advent of high-resolution instruments had brought us many new insights on the properties of MHD waves in coronal holes which are reviewed in this article. The advances made in the identification of compressive slow MHD waves in both polar and equatorial coronal holes, their possible connection with the recently discovered high-speed quasi-periodic upflows, their dissipation, and the detection of damping in Alfv\'{e}n waves from the spectral line width variation are discussed in particular.
\end{abstract}

\begin{article}

\section{Introduction}
Solar corona, when seen in ultraviolet (UV) and X-ray radiation, show dark patches on the Sun, called coronal holes. They appear dark only in the layers of the solar atmosphere with temperatures exceeding 10$^{5}$~K, making them indistinguishable from surroundings in the photospheric and low chromospheric images and hence the name `coronal' holes. These regions are cooler and less dense than their surroundings and are associated with open magnetic fields of the Sun. Magnetic flux is mainly unipolar with huge imbalance between the positive and negative fluxes \citep{2009SSRv..144..383W}. This is connected with the migration of decaying active region fields towards poles, and consequently, the location of coronal holes depends on solar cycle. During solar minimum, coronal holes are mostly confined to polar regions (polar coronal holes) while they can be found at  lower latitudes (equatorial coronal holes) as well, during solar maximum. The fast component of the solar wind emanates from the coronal hole regions. Hence, their presence influences the speed of slow solar wind coming from equatorial regions which reaches up to 600~km~s$^{-1}$ \citep{2003JGRA..108.1144Z} during solar maximum, faster than the average speeds ($\approx$400~km~s$^{-1}$) \citep{1997GeoRL..24.2885W}. Interested readers are referred to an excellent review by \citet{2009LRSP....6....3C}, for further information on coronal holes.

Polar coronal holes are large and persistent for most part of the solar cycle. Solar wind emanating from these regions is fast ($\approx$750~km~s$^{-1}$), tenuous and relatively homogeneous compared to the slower, denser and highly variable wind emanating from low latitude regions \citep{1997GeoRL..24.2885W, 2000JGR...10510419M}. During solar minimum, the global structure of solar wind is more or less uniform with the highly variable slow wind restricted to a narrow belt around equatorial regions whereas during solar maximum, it becomes complex with flows arising from multiple sources (streamers, active regions etc.,) at all latitudes \citep{2002GeoRL..29.1290M,2008GeoRL..3518103M}.

One of the striking features visible in the polar coronal holes is bright ray like structures above the limb called polar plumes that were first noticed in white light eclipse observations \citep{1950BAN....11..150V, 1958PASJ...10...49S, 1965PASJ...17....1S}. With predominantly unipolar magnetic flux concentrations at their foot points \citep{1997SoPh..175..393D}, plumes are believed to trace open magnetic fields in solar corona \citep{1998A&A...337..940B, 1999A&A...350..286Y}. The dark regions between plumes are called interplume regions. Plumes are cooler but denser than the surrounding interplume regions \citep{2006A&A...455..697W}. It is often debated whether plumes or interplumes are preferred sites for the acceleration of fast solar wind. While some researchers \citep{1988Sci...241.1781W, 2003ApJ...589..623G, 2005ApJ...635L.185G} believe plumes to be a major source of high speed wind, magnetic flux observations at the foot points \citep{1997ApJ...484L..75W} combined with lower outflow speeds \citep{2000A&A...353..749W, 2000A&A...359L...1P} and narrower spectral profiles in plumes \citep{1997ASPC..118..273A, 2000SoPh..194...43B, 2003ApJ...588..566T}  tend to support otherwise. The reason for this ambiguity can be partly attributed to our lack of knowledge on the exact physical process behind the acceleration of fast solar wind. Magnetohydrodynamic (MHD) waves are believed to be one of the possible candidates for transporting and depositing the required energy for acceleration. In coronal holes, plumes are regions of  Alfv\'{e}n speed minima (a consequence of the pressure balance across these dense structures) which makes them natural guides for different MHD waves \citep{2006RSPTA.364..473N}. Observations of these waves had been revolutionized by the launch of  Solar and Heliospheric Observatory \citep[SOHO,][]{1995SoPh..162....1D} in 1995. Since then, MHD waves were detected by several authors both from remote sensing and {\it in situ} observations. \citet{2007SoPh..246....3B, 2011SSRv..158..267B} provide a comprehensive review of these observations. Theoretical models were also developed  to understand the actual energy transfer process by these waves \citep[see the review by][]{2005SSRv..120...67O}.

Observations of wave activity, relatively simple geometry, and their connection to the fast solar wind, makes coronal holes interesting to study. In this article, we review the observations of MHD waves in coronal holes focusing primarily on progress made in the past few years. We also discuss on the new evidences of wave damping and highlight how such observations can be used to probe coronal conditions through seismology. Seismology is an interesting application of observing waves in the solar atmosphere but the subject is beyond the scope of the present review.
\section{Waves in Polar Coronal Holes}
During solar minimum, coronal holes are mostly confined to the polar regions. The off-limb part of the polar coronal holes appear structured with the presence of plume and interplume regions while the on-disk part is mostly dark with a few small bright patches (coronal bright points) appearing here and there. Numerous observations using imaging and spectroscopic techniques, revealed the presence of different MHD waves in these structures that can be categorized into compressive and incompressive waves.
\subsection{Compressive Waves} 
Compressive waves, as the name suggests, are perturbations that cause fluctuations in density, which thereby alter the observed emission or intensity. First such observations in polar plumes were made by \citet{1983SoPh...89...77W} who found short period (a few minutes) variations in Mg~{\sc x} emission observed by the Harvard Skylab experiment. However, the author did not find any corresponding variations in O~{\sc vi} line which is equally density sensitive and it was suggested that if these variations in Mg~{\sc x} emission are real they could be due to temperature fluctuations in response to the variations in local plasma heating rate. After the launch of SOHO, \citet{1997ApJ...491L.111O} observed fluctuations in polarized brightness (a measure of electron density) over a coronal hole high above the limb between 1.9 and 2.45~$R_{\odot}$, using data obtained with the white light channel of Ultraviolet Coronagraph Spectrometer (UVCS). This was the first undoubted detection of compressive waves in polar coronal holes. \citet{1998ApJ...501L.217D} studied the time evolution of several plumes by constructing evolution charts from a sequence of images observed by Extreme ultraviolet Imaging Telescope \citep[EIT,][]{1995SoPh..162..291D} onboard SOHO. Evolution charts were made by taking a strip along a plume at each instant and stacking them in time. They found diagonal stripes in these charts representing the propagation of enhancements and suppressions in brightness and suggested them as compressive waves propagating along the plumes. The period at which the stripes are repeated is 10 -- 15 min and the propagation speeds obtained from the inclination of the stripes, were 75 -- 150 km~s$^{-1}$. The amplitudes were found to be 10 -- 20\% of the overall intensity. Further analysis indicates an increase in wave amplitude with height. In a gravitationally stratified medium, the falling density will naturally cause an increase in wave amplitude if the energy flux carried by the waves is conserved. The observed increase is in agreement with the MHD simulations for propagating slow waves leading to the interpretation of these waves as slow magneto-acoustic waves \citep{1999ApJ...514..441O}. A number of studies followed reporting such oscillations in plumes, interplumes, and coronal holes, using spectroscopic data obtained with Coronal Diagnostic Spectrometer (CDS) and Solar Ultraviolet Measurements of Emitted Radiation (SUMER) onboard SOHO and  Extreme ultraviolet Imaging Spectrometer (EIS) onboard {\it Hinode} \citep{2000SoPh..196...63B, 2001A&A...377..691B,  2001A&A...380L..39B, 2006A&A...452.1059O, 2007A&A...463..713O,  2009A&A...499L..29B, 2009A&A...493..251G}. 

\begin{figure}
\centering
 \noindent\includegraphics[width=20pc]{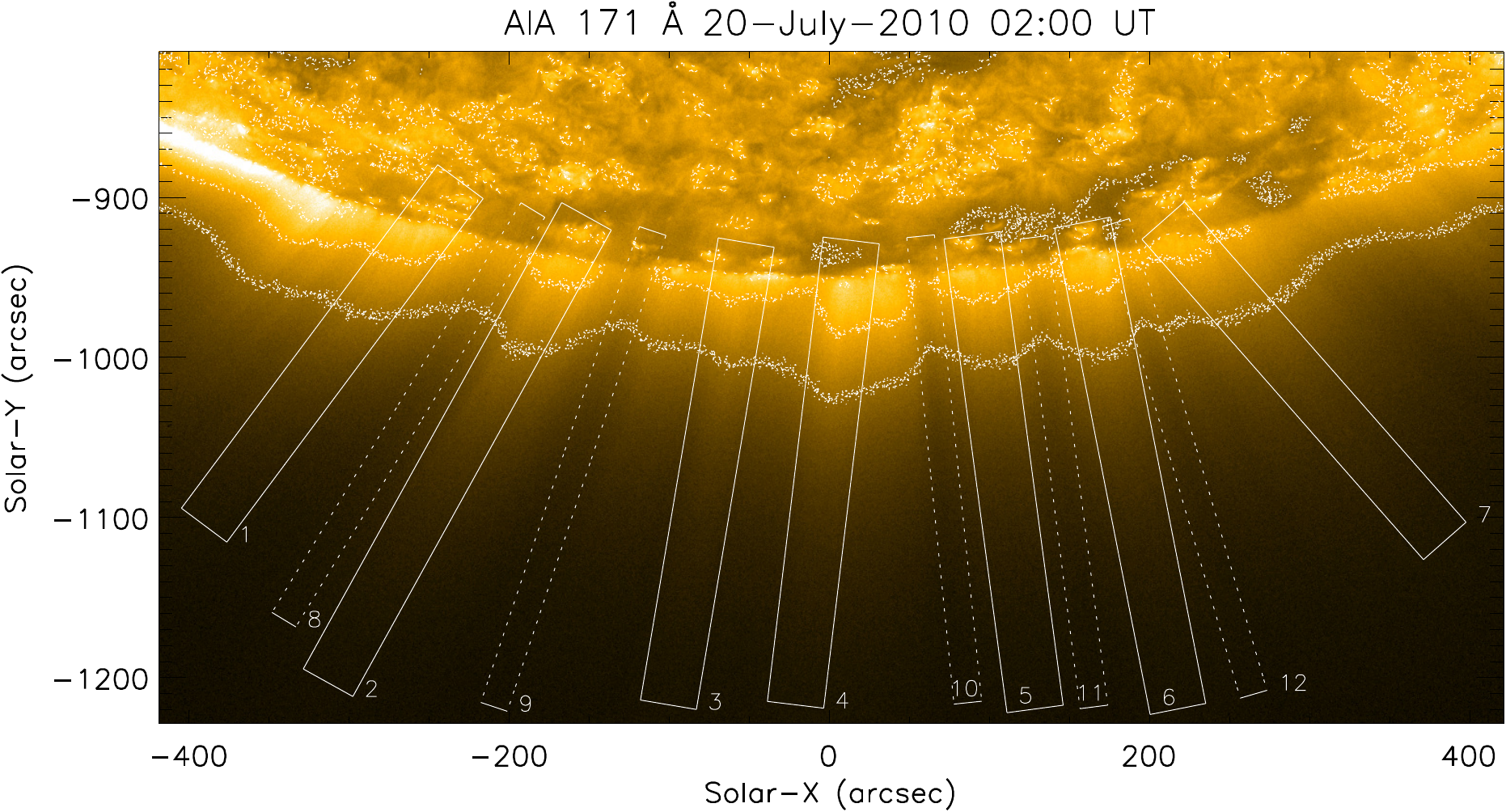}\\
\hspace*{0.13in}
 \noindent\includegraphics[angle=90,width=9.5pc]{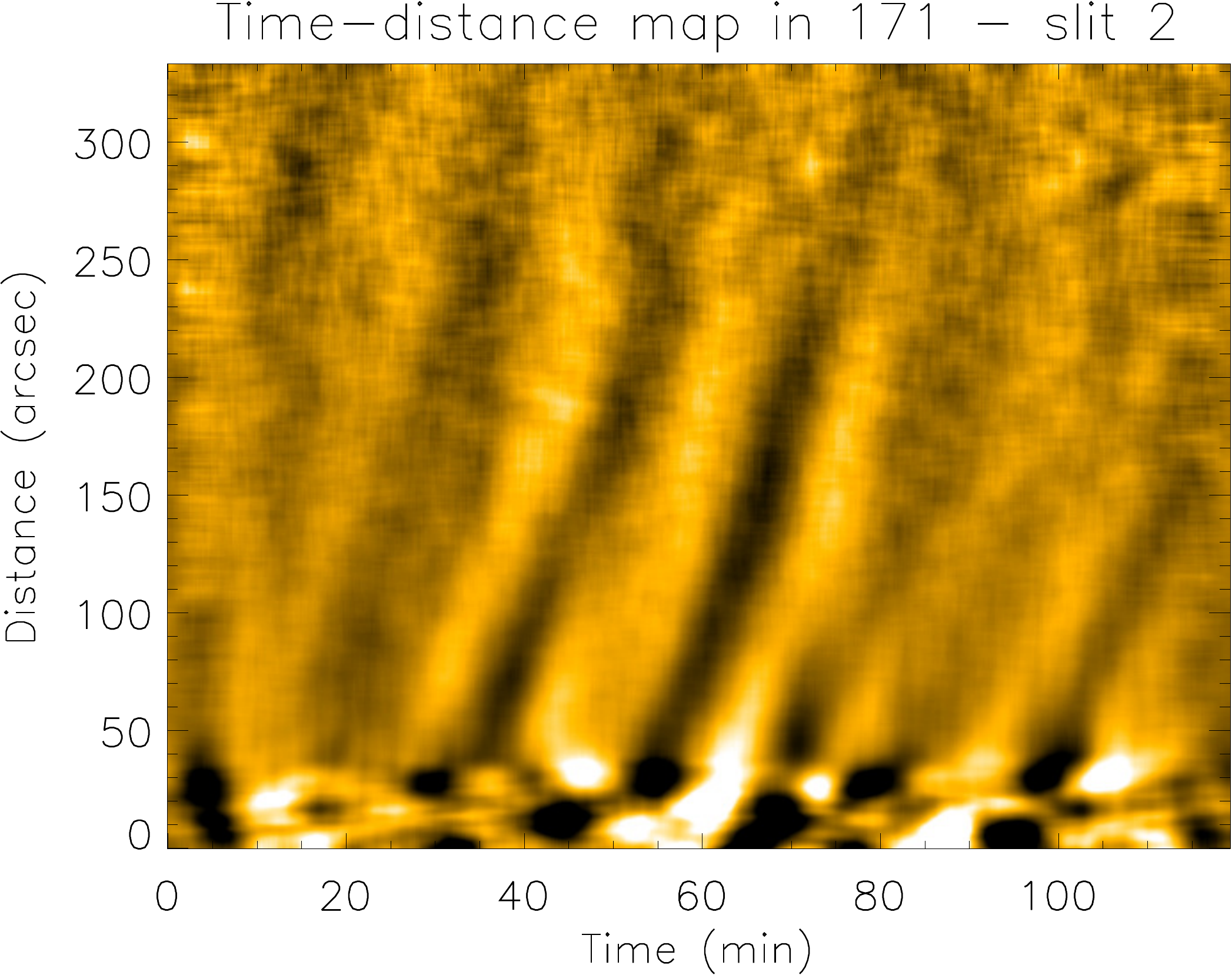}
 \noindent\includegraphics[angle=90,width=9.5pc]{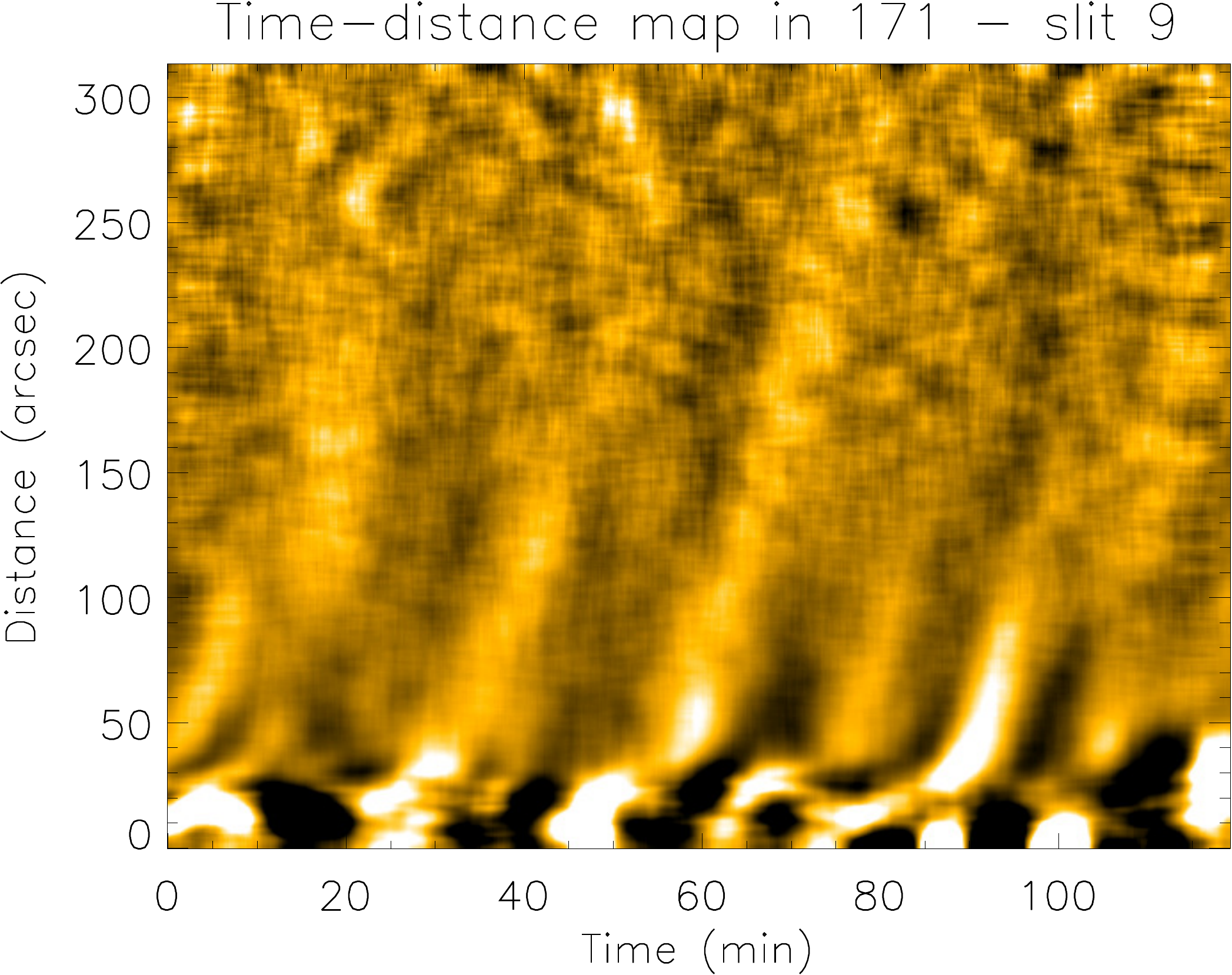}
 \caption{{\it Top:} Locations of all the slits chosen over several plume/interplume regions seen at a south polar region. Contours of different intensity levels overplotted as dotted curves, show the exact locations of plumes and interplumes. Slits over plume regions are 36$^{\prime\prime}$ wide and those over interplume regions are 18$^{\prime\prime}$ wide. {\it Bottom:} Time-distance maps constructed from slits 2 \& 9 as annotated. Adapted from \citet{2011A&A...528L...4K}.}
 \label{fig1}
\end{figure}

Using Extreme UltraViolet (EUV) observations from Solar TErrestrial RElations Observatory (STEREO), \citet{2010A&A...510L...2M} analysed several plume structures and suggested that propagating disturbances (PDs) along polar plumes could be due to the collimated high-speed quasi-periodic plasma jets that have similar properties rather than due to slow magneto-acoustic waves. Further, they conjectured that these jets could be responsible for loading a significant amount of heated plasma into the fast solar wind. It is important to clearly understand the nature of these PDs to find their contribution (mass/energy loading) to the solar wind. In an attempt to resolve this \citet{2011A&A...528L...4K} studied the properties of PDs along several plume and interplume regions using data obtained with Atmospheric Imaging Assembly (AIA) onboard Solar Dynamics Observatory (SDO). The authors followed the regular time-distance analysis, however, they chose wider (several tens of AIA pixels) slits along plume/interplume regions and averaged across them while constructing the time-distance maps. Figure~\ref{fig1} displays a south polar region with chosen slits over several plume and interplume regions marked (top panel). The width of the slits shown are 60 AIA pixels ($\approx$ 36$^{\prime\prime}$) for plume region and 30 AIA pixels ($\approx$ 18$^{\prime\prime}$) for interplume region. The bottom two panels of the figure show time-distance maps corresponding to slit 2 (plume) and slit 9 (interplume). PDs can be clearly seen in these maps and were found to be unaffected by variations in the  slit width. The faint plasma jets which occur at random should not be present in these maps as the spatial average across the slit makes them fainter. On the other hand, the stronger jets were identified visually and eliminated. The fact that the observed PDs are unaffected by variations in slit width implies a coherent behaviour. The projected propagation speeds were in the range of 100 -- 170 km~s$^{-1}$ and were found to be temperature dependent. Also, the speeds were slightly higher in the interplume regions possibly due to relatively hotter temperatures. Based on these properties, the authors infer that the observed PDs are due to slow MHD waves. Further, this study shows a way to avoid the contamination from plasma jets by using wider slits.

\begin{figure}
\centering
 \noindent\includegraphics[width=20pc]{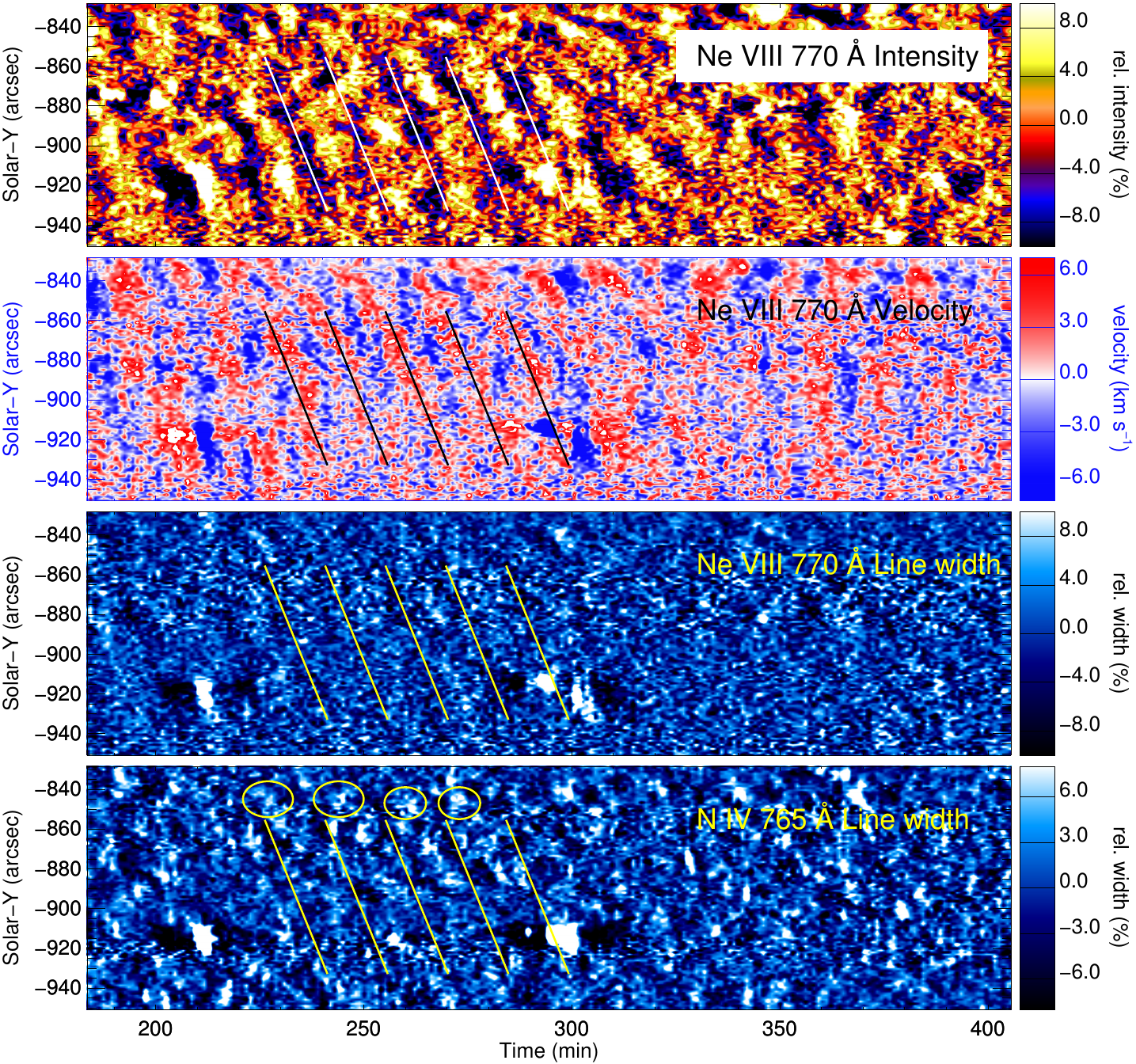}\\
 \caption{Enhanced time-distance maps for all the parameters of Ne~{\sc viii} line and for line width of N~{\sc iv} line as annotated. The slanted solid lines marked along the PDs indicate a periodicity of 14.5 min and propagation speed of 60 km~s$^{-1}$. Ellipses in the bottom panel show periodic enhancements in line width at the base of the PDs. From \citet{2012A&A...546A..93G}.}
 \label{fig2}
\end{figure}
\citet{2012A&A...546A..93G} observed PDs in an on-disk part of a south polar coronal hole using spectroscopic data obtained with SUMER. The spectral lines Ne~{\sc viii} 770~\AA, and N~{\sc iv} 765~\AA, which form at lower corona and transition region heights respectively, were used in this analysis. Time-distance maps for Ne~{\sc viii} show clear signatures of PDs in intensity and Doppler shift but are barely visible in line width. Similar maps for N~{\sc iv} line did not show any propagating feature in any of the line parameters. However, the map corresponding to line width shows periodic enhancements at the bottom of the PDs observed in Ne~{\sc viii} line. Figure~\ref{fig2} displays the time-distance maps of all the three line parameters for Ne~{\sc viii} line and that of line width for N~{\sc iv} line. The slanted solid lines indicate a propagation speed of 60 $\pm$ 4.8~km~s$^{-1}$ and a periodicity of 14.5 min. The ellipses in the bottom panel mark the periodic increase in line width. Note the foot point is at the top with limb towards the bottom so the slanted ridges indicate PDs propagating outward. This was the first time PDs were simultaneously observed in intensity and Doppler shift in a polar coronal hole. The enhancements in intensity correspond to a blue shift in velocity which is expected for an outward propagating slow magneto-acoustic wave. Further, line profiles were found to be pretty symmetric with very small asymmetries that were shown to arise from the finite sampling and noise in the data. The authors conclude that the line width enhancements at the bottom of the PDs could be possibly due to some unresolved explosive events while the PDs at greater heights show dominant wave signature similar to the case observed by \citet{2011ApJ...737L..43N} for an active region. A 3D MHD model developed by \citet{2012ApJ...754..111O} for a bipolar active region, also supports these results.

\citet{2012ASPC..456...91W} studied PDs in coronal loops using Hinode/EIS observations and forward models. They analysed the excess emission in the line profiles by subtracting a background profile instead of the regularly used Red-Blue (RB) asymmetry analysis method. Their results favor the wave interpretation for the observed PDs. From a statistical study, \citet{2012ApJ...759..144T} found two types of oscillations, one close to the loop foot points which show coherent oscillations in all the line parameters with blueward asymmetries in the line profiles, and the other associated with upper parts of the loops which show prominent oscillations in Doppler shift, but no significant variation in intensity and line width and no obvious asymmetries in the line profiles. The authors interpret the former type as due to episodic high-speed upflows and the latter type as kink/Alfv\'{e}n waves. \citet{2013ApJ...775L..23W} performed 3-D MHD simulations of PDs in long fan loops, and modeled a section of a loop with open boundary conditions at the top. They injected repetitive tiny flow pulses at the base of the coronal loop, with an energy frequency distribution that follows the flare power-law scaling. It was found that the injected upflow pulses inevitably excite slow magneto-acoustic waves in the loop. The upflows were found to rapidly decelerate with height and the resulting PDs are mostly dominated by the wave signatures except near the base where the contribution from flows can be notable. Based on these results, the authors suggested that nanoflare-like impulsive heating events at the base of the loop may produce both upflows and waves.

Following a different approach, \citet{2014ApJ...790..150S} studied the nature of PDs observed at both north and south polar regions using 21 datasets. By fitting an exponentially growing sine curve to the spatial variation of PDs very close to the limb (0--9~Mm), they measured the pressure scale height and wavelength in three AIA channels 171~\AA, 193~\AA, and 211~\AA. The temporal variation was then used to estimate the oscillation period in these channels. Average values of scale height obtained for all the sets of data are 6.4$\pm$2.8, 4.8$\pm$3.2, and 3.3$\pm$1.7~Mm, respectively in 171~\AA, 193~\AA, and 211~\AA\ channels. Corresponding wavelength values are 35$\pm$7, 26$\pm$11, 25$\pm$20~Mm. The periodicity on the other hand, is roughly same ($\approx$ 24 min) in all three channels. Although the uncertainties in the scale height and wavelength are large, the authors had statistically shown that they are actually temperature dependent wheras the periodicity remains independent of temperature. Further, they calculated the propagation speeds from the ratios of wavelength to time period and compared those with the independent estimates obtained from the pressure scale height (through its relation to temperature). A good agreement was found between these values. Depending on these properties, the authors suggest that the detected PDs are slow magneto-acoustic waves in nature. However, the opposite (to expected) dependence of these parameters on temperature, i.e., shorter scale heights and wavelengths in hotter channels, is not explained by the authors.

 \citep{2010ApJ...718...11G} found accelerating PDs in interplume regions. The authors combined spectroscopic observations from SUMER and 40$^{\prime\prime}$ slot images from EIS to study the properties of PDs in plume and interplume regions. Their analysis reveals PDs of 15-20 min periodicity in interplume regions travelling with projected speeds increasing from 25$\pm$1.3~km~s$^{-1}$ at their origin to 330$\pm$140~km~s$^{-1}$ at around 160$^{\prime\prime}$ above the limb. These speeds were also found to be temperature independent. PDs of periodicity in the same range were found in the adjacent plume region but with no significant acceleration. Although the observations were mainly in intensity, the acceleration and other properties (such as the supersonic speeds and their temperature independence) led the authors to conclude the PDs observed in interplume region as due to either Alfv\'{e}nic or fast magneto-acoustic waves and those in the plume region as due to slow magneto-acoustic waves.

Slow magneto-acoustic waves were also observed in the on-disk part of a coronal hole \citep{2007A&A...463..713O,2009A&A...493..251G}. Spectroscopic observations in N~{\sc iv} 765~\AA (transition region) and Ne~{\sc viii} 770~\AA (low corona) lines, obtained with SUMER, were used by \citet{2009A&A...493..251G} to study the phase delays between intensity and velocity oscillations. They found upward and downward propagating waves in the internetwork regions and only upward propagating waves in the network regions of the polar coronal hole.

\subsection{Incompressive Waves}
Incompressive waves do not cause any perturbations in intensity, instead they alter other spectral parameters. Therefore, spectroscopic observations are necessary to observe them. For instance, the transverse Alfv\'{e}n waves which are incompressible in the linear limit, cause oscillations only in Doppler shift. However, the emission in polar coronal holes, being low, often require longer exposures, spatial, and temporal averaging to obtain a good signal. This makes the Alfv\'{e}n waves unresolved (if the time period is shorter than the temporal bin) and increases Doppler width rather than causing a Doppler shift. In the WKB approximation, the Alfv\'{e}n wave energy flux ($F$) crossing a surface of area $A$, can be written as 
\begin{equation}
F = \rho \langle \delta v^2\rangle  v_A A =  \sqrt\frac{\rho}{4\pi}\langle \delta v^2\rangle B A
\label{eq1}
 \end{equation}
where $v_{A}= B/\sqrt{4\pi \rho}$ is Alfv\'{e}n wave velocity, $\rho$ is plasma mass density, $B$ is field strength, and $\langle\delta v^{2}\rangle$ is the mean square velocity related to the observed non-thermal velocity $\xi$ as $\xi^{2}=\langle\delta v^{2}\rangle/2$. It may be noted that this approximation is not well applicable when wave reflections or refractions are strong, and when damping is important. Nevertheless, it is useful to describe the general propagation. For a flux tube geometry $B A$ remains constant, and if the energy flux is conserved (i.e., if the Alfv\'{e}n waves are undamped as they propagate), then the non-thermal velocity $\xi$ follows
\begin{equation}
 \xi \propto \rho^{-1/4}.
\label{eq2}
\end{equation}

Equation~\ref{eq2} implies, the observed Doppler width due to (unresolved) Alfv\'{e}n waves increases with fall in density as they propagate outwards. This particular feature turned out to be an important (although indirect) evidence for the presence of Alfv\'{e}n waves. Several authors observed an increase of line width with height in polar coronal holes suggesting the interpretation of outward propagating undamped Alfv\'{e}n waves \citep{1994SSRv...70..373H, 1998A&A...339..208B, 2004A&A...415.1133W, 2009A&A...501L..15B}. A few others observed an initial increase followed by a decrease or level off at greater heights \citep{2003A&A...400.1065O, 2005A&A...436L..35O, 2008A&A...480..509M} and indicated that the decrease in line width could be due to Alfv\'{e}n wave damping. 

\begin{figure}
\centering
 \noindent\includegraphics[width=20pc]{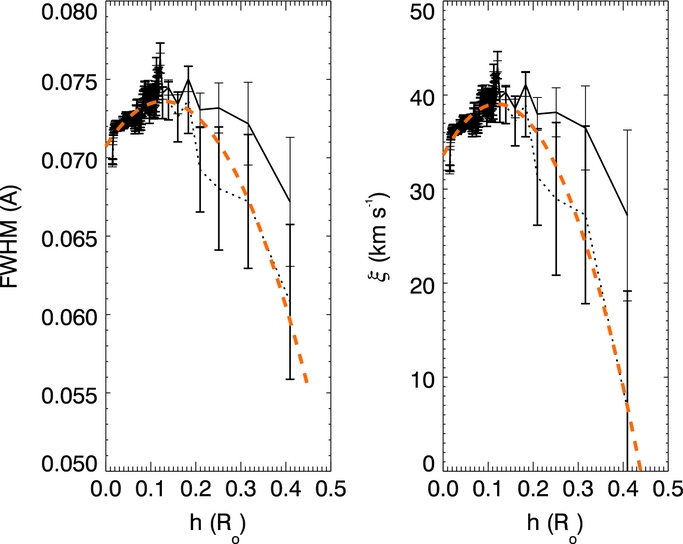}\\
 \noindent\includegraphics[width=20pc]{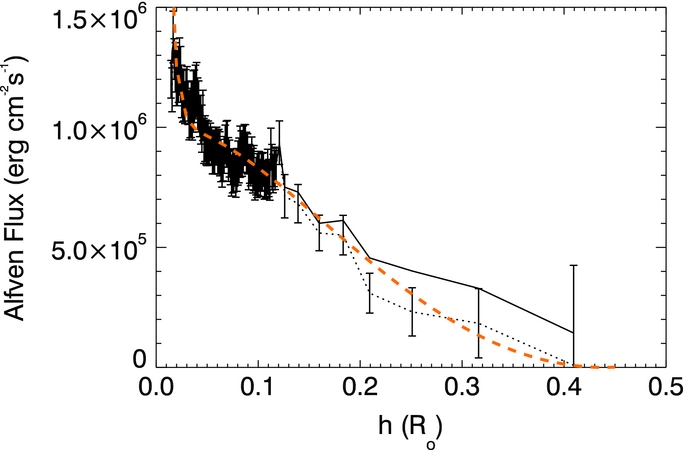}
 \caption{{\it Top:} Variation of line width (left) and non-thermal velocity (right) with height above the limb for the spectral line Fe~{\sc xii} 195.12~\AA. Solid and dotted lines represent values before and after the subtraction of Fe~{\sc xii} stray light profile, respectively. Dashed lines represent parabolic fits to the corrected data. Note the non-thermal velocities shown here are devoid of instrumental broadening but the line width values are not. {\it Bottom:} Alfv\'{e}n wave energy flux as a function of height above the limb. Values shown in different line styles correspond to different non-thermal velocities shown in the top panel. The calculations were done for a constant magnetic field, $B$=8~G. Adapted from \citet{2012ApJ...751..110B}.}
 \label{fig3}
\end{figure}
\citet{2005A&A...436L..35O} also observed a turnover in the line width variation and found a change in excitation mechanism from collisional to radiative at the turning point. \citet{2003ApJ...585..516S} found increase and decrease of line widths in different lines of a simultaneously observed pair and suggested that the variation in line width may always not be related to propagating waves. \citet{2008A&A...483..271D} found that the decrease in line width is primarily due to scattered light that was not properly estimated. Further, the increase in line widths can also be caused by increase in ion temperatures \citep{1997ApJ...484L..87S, 1998ApJ...503..475T}. All these studies made the detection of Alfv\'{e}n waves in coronal holes inconclusive.

Recently, \citet{2012ApJ...751..110B} studied the variation of line width in a polar coronal hole using spectroscopic data from EIS. The observations were made with 1$^{\prime\prime}$ slit in sit-and-stare mode for nearly 21.6 hrs. The exposure times were kept long (500~s) to gather the signal up to farther off-limb distances. After careful subtraction of the stray light estimated using a transition region line, the authors found that the width of the Fe~{\sc xii} 195.12 \AA\ line increases up to $\simeq$0.14~$R_{\odot}$ and decreases higher up. Assuming ionization equilibrium, they also calculated non-thermal velocity as a function of height which shows similar behaviour. Figure~\ref{fig3} displays the variation of line width and non-thermal velocity with height above the limb for Fe~{\sc xii} 195.12~\AA\ line (top panel). The solid and dotted lines in the plots correspond to the results before and after the subtraction of additional stray light caused by a blend in the Fe~{\sc xii} line, respectively. Dashed lines represent parabolic fits to the corrected data. The line width values shown here, were not corrected for the instrumental broadening, although the correction was done before the computation of non-thermal velocities. These plots clearly show the decrease of non-thermal velocity after an initial increase. It should be noted that the non-thermal velocity is computed here by assuming a constant ion temperature at all altitudes, so any possible increase in ion temperature with height will further reduce the non-thermal velocities making the decrease even steeper. The authors interpreted this behaviour as due to damping in Alfv\'{e}n waves and estimated the energy flux assuming a constant magnetic field $B$=8~G. The computed values as a function of height were shown in the bottom panel of Figure~\ref{fig3}. Values shown in different line styles correspond to different non-thermal velocities plotted in the top panel. Apparently, the energy flux continuously decreases even in the low altitude region where the line width increases with height. It goes down from about $\approx$1.2$\times$10$^{6}$ erg cm$^{-2}$ s$^{-1}$ at 0.03~$R_{\odot}$ above the limb to about $\approx$8.5$\times$10$^{3}$ erg cm$^{-2}$ s$^{-1}$ at 0.4~$R_{\odot}$ above the limb. 
\begin{figure}
\centering
 \noindent\includegraphics[width=20pc]{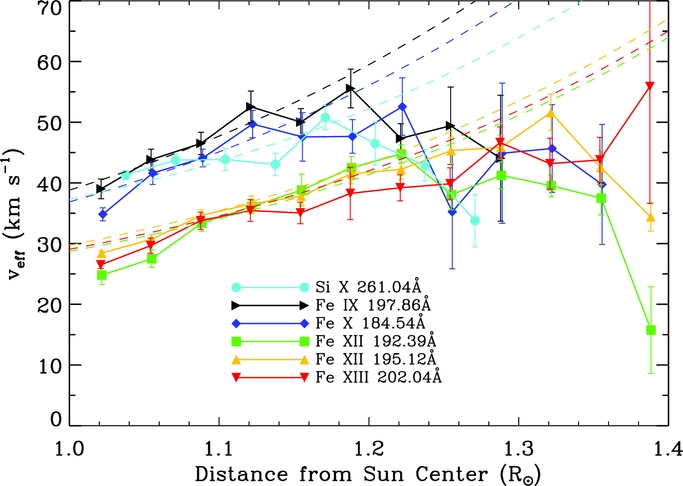}
 \caption{Variation of Doppler width (after instrumental correction) as a function of radial distance for several spectral lines as indicated in the plot. The dashed lines represent the corresponding values expected for undamped wave propagation. From \citet{2012ApJ...753...36H}.}
 \label{fig4}
\end{figure}
In an independent study, \citet{2012ApJ...753...36H} also found a decrease in line width with height following an initial increase in a polar coronal hole. The scattered light contribution was carefully removed and the authors rule out any effects due to line of sight and changes in excitation mechanism suggesting the Alfv\'{e}n wave damping. Figure~\ref{fig4} displays their results for several spectral lines. The turning point seems to be at different heights for different lines indicating multiple structures in the line of sight. Dashed lines in the figure mark the respective values for undamped wave propagation. Assuming thermodynamic equilibrium, the energy flux estimated was about $\approx$7.0$\times$10$^{5}$ erg cm$^{-2}$ s$^{-1}$ at 0.1~$R_{\odot}$ above the limb to about $\approx$1.1$\times$10$^{5}$ erg cm$^{-2}$ s$^{-1}$ at 0.3~$R_{\odot}$ above the limb. The dissipated energy between these heights amounts to 70\% of the energy required to heat the coronal hole and accelerate the fast solar wind ($\approx$8$\times$10$^{5}$ erg cm$^{-2}$ s$^{-1}$; \citet{1977ARA&A..15..363W}). However, the authors cautioned that the estimated values would be upper bounds since the actual ion temperatures can be higher than the electron temperatures used here for the estimation of non-thermal velocity. Although suffered by some observational uncertainties and involve certain assumptions, both these studies seem to provide a compelling evidence for the existence and damping of Alfv\'{e}n waves in coronal holes.

\section{Waves in Equatorial Coronal Holes}
Coronal holes are also observed in the equatorial regions during the maximum phase of the solar cycle. Equatorial coronal holes are usually associated with the remnant active regions. Their smaller spatial extent makes them easy to observe when they are at the disk center, which however, poses a challenge for observations of propagating waves expected for an open field configuration. Incompressive waves with perturbations perpendicular to the line of sight are difficult to observe, nevertheless, there are a few studies which report the presence of compressive waves.  \citet{2007A&A...463..713O} performed a statistical study to find evidence for waves in equatorial coronal holes using observations from CDS/SOHO. Oscillations in intensity and Doppler shift were studied for the spectral lines O~{\sc v} 629~\AA, Mg~{\sc x} 624~\AA, and Si~{\sc xii} 520~\AA. By measuring phase delays between different line pairs, propagation speeds were  estimated to be in the range of 50 to 120 km~s$^{-1}$. It was also observed that the oscillations occur preferentially in bright regions that are associated with magnetic field concentrations. The authors interpreted the observed oscillations as primarily due to propagating slow magneto-acoustic waves. The energy flux carried by these waves was estimated to be $\approx$ 2.5$\times$10$^{4}$ erg cm$^{-2}$ s$^{-1}$ using WKB approximation, which is lower than that required for a coronal hole with high speed wind ($\approx$ 8$\times$10$^{5}$ erg cm$^{-2}$ s$^{-1}$). 

\citet{2008A&A...488..331T} used observations from Transition Region and Coronal Explorer (TRACE) to study intensity oscillations in a network region at the boundary of an equatorial coronal hole. Images from two different passbands 1600 \AA\ and 171 \AA\  representative of chromosphere and low corona, respectively, were used in the analysis. Power maps constructed in three frequency regimes, high (1.2 -- 2.0 mHz), intermediate (2.6 -- 4.0 mHz), and low (5.0 -- 8.3 mHz), indicate a suppression of high frequency power over the magnetic network in both the passbands suggesting the extension of magnetic shadows to lower corona. Detailed investigation on individual features showed PDs of 5-10 min periodicity, propagating at speeds of the order of the coronal sound speed. Interpreting the observed PDs as slow magneto-acoustic waves, the authors calculated the energy flux contained in these waves corresponding to both the passbands. For 171~\AA\ the value is about 40 erg cm$^{-2}$ s$^{-1}$, which is far below the energy requirement for quiet corona and that for 1600 \AA, is about 1.368$\times$10$^{6}$ erg cm$^{-2}$ s$^{-1}$, which is of the order of the required energy for chromosphere.

\begin{figure*}
\centering
 \noindent\includegraphics[angle=90,width=30pc]{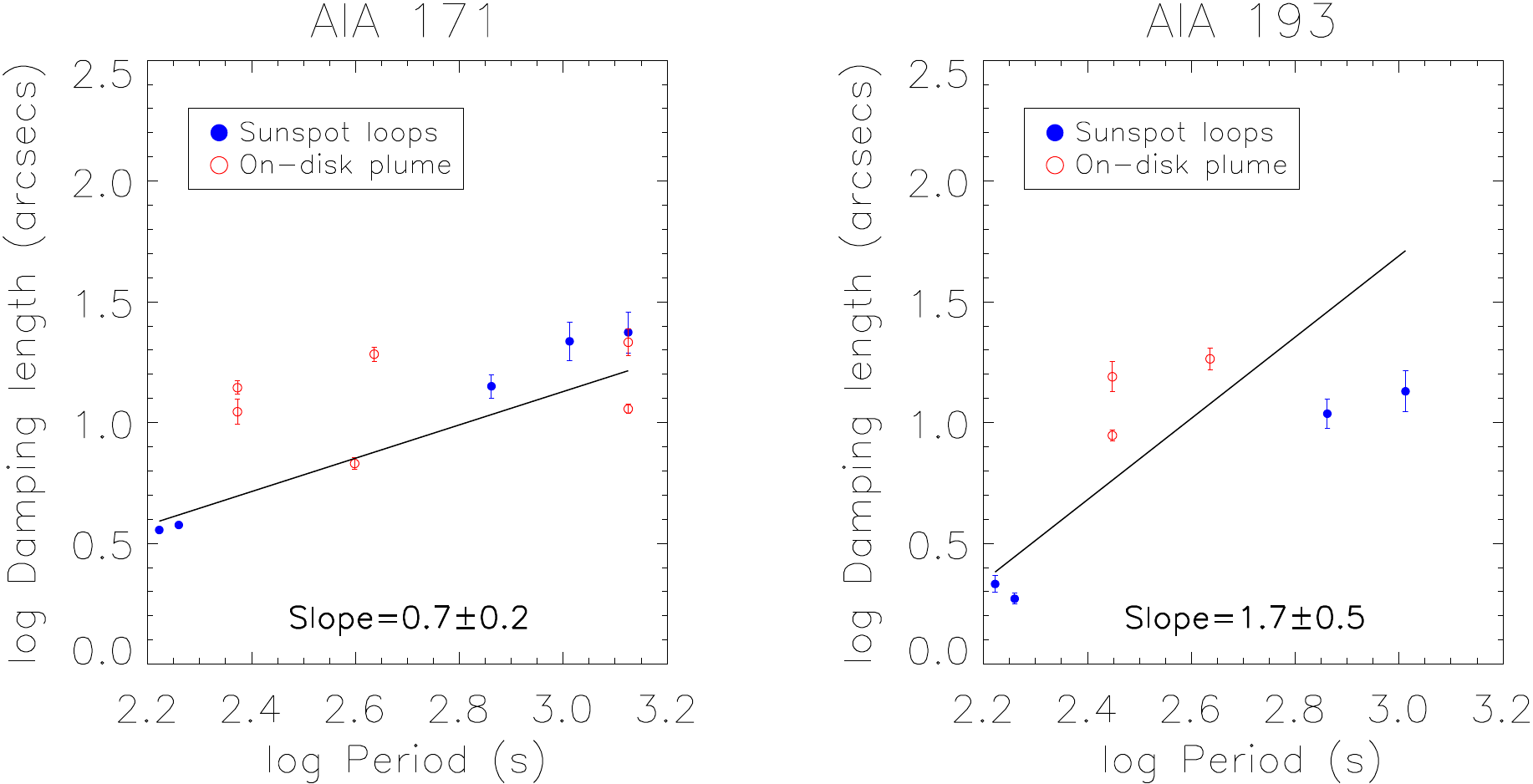} \\
 \noindent\includegraphics[angle=90,width=30pc]{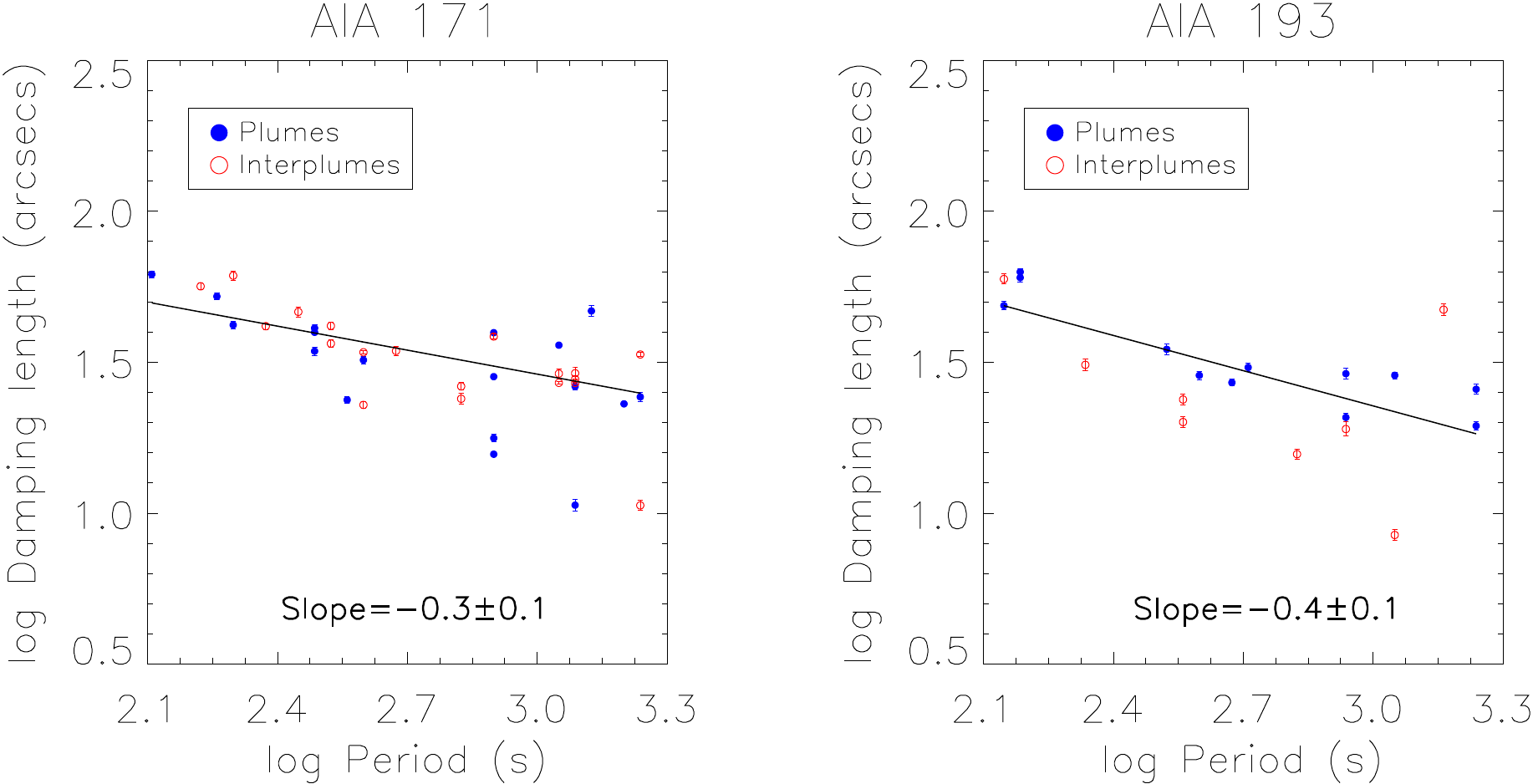}
 \caption{Dependence of observed damping length on oscillation period for slow magneto-acoustic waves. Panels on the top display results for structures on-disk and those at the bottom display results for polar regions above the limb. Left and right panels are for 171 \AA\  and 193 \AA\ channels of AIA, respectively. Adapted from \citet{2014ApJ...789..118K}.}
 \label{fig5}
\end{figure*}

\begin{table*}
\begin{center}
\caption{Dependence of damping length on oscillation period (P ) for slow waves damped by different physical mechanisms. From \citet{2014ApJ...789..118K} }
\label{tab1}
\begin{tabular}{c c c}
\hline\hline
 Physical mechanism & Amplitude growth of  & Period dependence of \\
                    & density perturbations & damping length ($L_{d} \propto$) \\
 \hline
 Thermal conduction  &  & \\
  lower limit  & $e^{-\frac{d\omega^2}{2c_{s}}(\gamma-1)z}$  & $P^{2}$ \\
  upper limit  & $e^{-\frac{(\gamma-1)z}{2d\gamma^{3/2}c_{s}}}$  & $P^{0}$ \\
 Compressive viscosity & $e^{-\frac{2}{3}\frac{\eta_{0}\omega^2z}{\rho_{0}c_{s}^3}}$  & $P^{2}$ \\
 Optically thin radiation& $e^{-\frac{r_{p}z}{c_{s}}}$  & $P^{0}$ \\
 Gravitational stratification & $e^{\frac{-z}{2H}}$  & $P^{0}$ \\
 Magnetic field divergence & $R_{\odot}/r$  & $P^{0}$ \\
\hline
\end{tabular}
\end{center}
\end{table*}
\section{Dissipation/Damping in waves}
One of the most desirable characteristics of MHD wave observations in the solar atmosphere, is their dissipation. The  energy carried by these waves has to be deposited at appropriate heights to facilitate coronal heating and solar wind acceleration. It turned out that the compressive waves are easy to dissipate with the conventional physical mechanisms whereas the incompressive waves require some special conditions.

Propagating slow magneto-acoustic waves were known to disappear after travelling some distance along the guiding structure. Their amplitude decays rapidly as they propagate outwards. Results from theoretical modelling suggest that thermal conduction, compressive viscosity, optically thin radiation, area divergence, and gravitational stratification affect the wave amplitude \citep[][and the references therein]{2009SSRv..149...65D}. Recent observations by \citet{2012A&A...546A..50K} reveal that the damping in these waves is frequency-dependent. The authors constructed power maps in different periodicity ranges and found that the power in long-period range is significant up to longer distances compared to that in the short-period range. This dependence, evidently, comes from the damping mechanism and hence useful to identify it. To further explore this, \citet{2014ApJ...789..118K} studied the quantitative dependence of damping length on oscillation period in several on-disk and off-limb structures. Figure~\ref{fig5} displays their results for sunspot loops and on-disk plume in the top panels and that for plume interplume regions above the limb in the bottom panel. The left and right panels are for two different temperature channels of AIA (171~\AA\ and 193~\AA). Clearly, the dependence of damping length in polar regions is different from that in the on-disk regions. Besides, neither of them seem to agree with the expected theoretical dependence for any of the dominant physical mechanisms thought to be responsible for damping in slow waves. Table~\ref{tab1} lists the expected theoretical dependence of damping length on oscillation period for several physical mechanisms. These relations were derived following a 1D linear MHD model \citep{2003A&A...408..755D, 2004A&A...415..705D} that is applicable under the assumptions that the amplitude of the oscillations is small and plasma-$\beta$ is unity. This discrepancy between the theory and observations implies the current 1D linear MHD models are inadequate in describing the damping in slow waves. It is possible that the waves in polar regions undergo non-linear steepening that causes enhanced viscous dissipation \citep{2000ApJ...533.1071O, 2015A&A...573A..32A} which is not accounted for in this model. However, the observations did not indicate any signatures of non-linearity or shock formation. So it remains to be seen if phase mixing, mode coupling or some new mechanism plays a key role in slow wave damping. It may also be noted that more data is required particularly from the on-disk loops and plume-like structures to establish the exact dependence of damping length on the oscillation period.

\citet{2014A&A...568A..96G} also studied the dissipation of slow MHD waves in a polar coronal hole above the limb. The author found two different dissipation regimes, one close to the limb ($<$ 10 Mm) where the waves are heavily damped and the other at greater heights (10 -- 70 Mm) where the waves damp slowly. Frequency-dependence was also investigated in both these regimes by taking three period ranges 4--6 min, 6--15min, and 16--45 min. The dissipation at greater heights was found to be proportional to the period while that close to the limb do not show any preferred relation. It was suggested that the dissipation at greater heights could be due to thermal conduction but the cause for heavy damping close to the limb is unclear. A direct comparison of this study with \citet{2014ApJ...789..118K} would not be possible since the results here correspond to three pre-defined frequency ranges whereas those from \citet{2014ApJ...789..118K} correspond to several individual frequency peaks identified from the Fourier power.

Dissipation of Afv\'{e}n waves by pure viscosity, resistivity, and thermal conductivity is not significant in a homogenous laminar equilibrium plasma at coronal temperature and density. However, the lower corona is not homogeneous, nor at equilibrium, and there is ample evidence of persistent fluctuations and turbulence in the coronal plasma that may affect the dissipation of the waves. In fact, if the plasma is inhomogeneous, processes like phase mixing, resonant absorption, nonlinear dissipation, and mode coupling, can enhance the dissipation of Alfv\'{e}n waves. For example, Alfv\'{e}n waves with opposite phases on neighboring field lines enhance the viscous and resistive dissipation rates \citep{1983A&A...117..220H}. In coronal holes, perhaps the most relevant disssipation mechanism for Alfv\'{e}n waves is via turbulent cascades. Low-frequency Alfv\'{e}n waves get reflected due to the gradients in magnetic field and density, and interact with the outward propagating waves causing a turbulent cascade that transfers energy to smaller scales. Upon reaching sufficiently small scales, the energy is dissipated to the ambient medium through wave-particle interactions. Several theoretical models based on this mechanism were proposed to explain the acceleration of fast solar wind in coronal holes \citep{1968ApJ...153..371C, 2004JGRA..109.7102O, 2009ApJ...707.1659C, 2010ApJ...708L.116V, 2014ApJ...796...43P}. Observationally the damping in Alfv\'{e}n waves has been detected indirectly from the variations in spectral line width \citep{2012ApJ...751..110B, 2012ApJ...753...36H}. While most of the previous studies suffered from scatter light and other instrumental problems these authors had overcome some of those difficulties and seem to show a clear decrease in line width with height, an indication of Alfv\'{e}n wave damping.

\section{Summary and Conclusions}
MHD waves are integral part of the coronal hole dynamics. The uninterrupted view of full disk of the Sun by SDO delivering ultra high-resolution images and high-resolution spectroscopic observations by Hinode led to a remarkable progress in our understanding on these waves in the past few years. High speed quasi-periodic upflows were observed in open structures and found to cause an ambiguity with the outward propagating slow magneto-acoustic waves. In polar plume/interplume regions this can be partly avoided by using the coherence nature of waves \citep{2011A&A...528L...4K} and it was found that the flow-like behaviour is dominant only near the base of the guiding structure \citep{2011ApJ...737L..43N, 2012A&A...546A..93G}. Nevertheless, it appears that these periodic upflows at the base and slow waves are interconnected. Recent 3-D MHD simulations by \citet{2012ApJ...754..111O} and \citet{2013ApJ...775L..23W}, for a coronal loop, reveal that injecting tiny upflow pulses at the bottom inevitably excites slow magneto-acoustic waves that propagate along the loop, confirming this. The authors also indicate that small-scale impulsive heating events such as nanoflares at the loop base can produce both the upflows and waves. Perhaps, this could also explain the origin of long period (few tens of minutes) slow waves in the solar atmosphere. 

Besides providing the additional momentum for the acceleration of solar wind, the dissipation of slow waves, if understood well, can provide us with a wealth of information about the coronal conditions (for e.g., transport coefficients) through MHD seismology. The abundant observations of these waves in active region loops, coupled with the recent advances in theoretical modelling gave us significant understanding on their dissipation. Still, there appears to be a discrepancy between the theoretical and observed dependence of damping on the oscillation period of these waves \citep{2014ApJ...789..118K}. The observed dependence in plume/interplume regions was also found to be considerably different from that in the on-disk loops despite the similarity in the waves' properties. This latter disagreement could possibly arise from the differences in physical conditions of these two regions.

The incompressive Alfv\'{e}n waves were believed to carry sufficient flux to meet the coronal hole energy budget, but been elusive so far. There were some indirect evidences from the variation of line broadening with height above the limb but most of them are crippled by the inability to separate thermal/non-thermal components of line width and contamination from scattered light had been another hindrance. Recent observations by \citet{2012ApJ...751..110B} and \citet{2012ApJ...753...36H} seem to overcome these difficulties. The authors found the width of several spectral lines initially increase up to certain height above the limb and then steadily decrease even after carefully accounting for the scattered light. Since the ion temperatures do not decrease with height, this had been a compelling evidence for the existence and damping of Alfv\'{e}n waves in the polar regions. The dissipated energy flux estimated by these authors also match with the requirement for coronal holes. However, these observations could not provide any information on the exact dissipation mechanism. The source of these oscillations also remains unclear. We hope the recently launched Interface Region Imaging Spectrograph (IRIS) and the upcoming solar missions will shed more light on this subject.

% ACKNOWLEDGMENTS
\begin{acknowledgments}
DB wishes to thank the AGU for generous support which enabled him to attend the Chapman conference. 
\end{acknowledgments}

\bibliographystyle{agu08notitle9}
\bibliography{kpref}

\end{article}

% Enter Figures and Tables here:

\end{document}